\documentclass[prb,twocolumn,aps,showpacs]{revtex4}
\usepackage{graphicx} 
\newcommand{\beq}{\begin{eqnarray}} 
\newcommand{\eeq}{\end{eqnarray}}

\begin{document} 
\title{Theory of the Magnetic Moment in Iron Pnictides} 
\author{Jiansheng Wu$^1$, Philip Phillips$^1$ and A. H. Castro Neto$^2$} 
 
\affiliation{$^1$Department of Physics, 
University of Illinois at Urbana-Champaign, 
1110 W. Green Street, Urbana IL 61801, U.S.A.} 
\affiliation{$^2$Department of Physics, Boston University, 590 
  Commonwealth Avenue, Boston, Ma. 02215}

\begin{abstract} 
We show that the combined effects of spin-orbit, monoclinic distortion, 
and p-d hybridization in tetrahedrally coordinated Fe in LaFeAsO 
invalidates the naive Hund's rule filling of the Fe d-levels. 
The two highest occupied levels have one electron each but as a result 
of differing p-d hybridizations, the upper level is more itinerant while 
electrons in the lower level are more localized.   The resulting
magnetic moment is highly anisotropic with an in-plane value of $0.25-0.35\mu_B$ per Fe and a z-projection of $0.06\mu_B$, both
of which are in agreement with experiment. 
\end{abstract} 
 
 
\maketitle 
 
\begin{figure} 
  \includegraphics[scale=0.3]{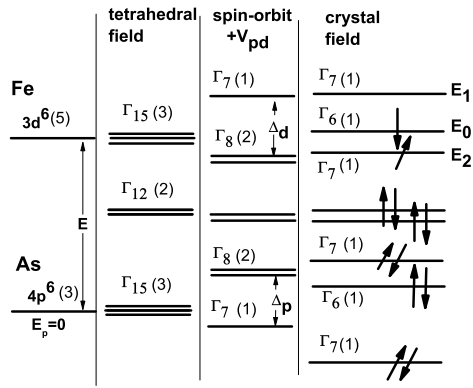}\\ 
  \caption{Evolution of the energy levels of the Fe 3d and As 4p 
     levels after the inclusion of spin-orbit coupling, p-d 
    hybridisation, $V_{\rm pd}$, and the monoclinic crystal field 
    distortion.} 
\label{fig1} 
\end{figure}

The representative parent material, LaFeAsO, in the rapidly growing class of iron-based 
superconductors\cite{s1,s3,s4,s5} exhibits a structural 
monoclinic distortion 
from tetragonal symmetry 
at $150K$ followed by a transition to an antiferromagnet\cite{lynn,o1,o2,o5} at 134K with a unit cell of ($\sqrt{2}\times\sqrt{2}\times 2)$. The 
observed magnetic moment per Fe atom has been reported to range from $0.25\mu_B$\cite{o5} to $0.36\mu_B$\cite{lynn} and lies in the 
a-b plane.  Such a low value of the magnetic moment is astounding 
because 
any application of Hund's rule to the Fe d-states 
results in a moment of at least $2\mu_B$.  We offer 
here a resolution of the low {\bf in-plane} magnetic moment in 
LaFeAsO which is rooted in three effects 
that are well known to be important 
in FeAs-based materials\cite{wolley,johansson,galanakis}, namely 
spin-orbit, strong hybridization between the Fe d and the As 
4p orbitals, and the lattice compression along the z-axis 
to the lower monoclinic symmetry. 
All three 
conspire to destroy the naive Hund's rule filling of the Fe 
atomic levels as illustrated in Fig.~\ref{fig1}.
 
That the properties of Fe-based materials are strong functions of 
the hybridization is not new.  Well known is the case of an isolated 
Fe atom which has a moment of 4$\mu_B$ whereas in the metal, the 
moment is roughly halved to 2.2$\mu_B$ per Fe as a result\cite{cheli} of the 4s 
and 3d hybridization.    Less 
well-known, but more pertinent to LaFeAsO, is the case of the 
Zincblende complex FeAs which is also an antiferromagnet and has a 
monoclinic distortion\cite{KA}. First principles calculations on Fe 
films deposited on GaAs \cite{johansson} reveal that the value of 
the magnetic moment per Fe is a strong function of Fe-As bond 
distance. The moment vanishes\cite{johansson} for Fe-As distances 
less than $2.36\AA$. This effect was attributed\cite{johansson} to 
the strong hybridization between the Fe 3d and the As 4p orbitals. 
In LaFeAsO, the average Fe-As bond distance, 2.4\AA, is close to the 
critical value of $2.36\AA$ found for Fe-As films.  As the degree to 
which Fe and As are non-coplanar in FeAs and LaoFeAs is identical, 
similar extreme sensitivity of the moment to the p-d  hybridization 
is expected in LaFeAsO. 
 
It is ultimately symmetry that dictates hybridization. In LaFeAsO, 
each Fe is tetrahedrally coordinated.  Full tetrahedral (cubic) symmetry 
splits the d-states into two irreducible representations\cite{wolley,galanakis}: 1) the three-fold 
degenerate $\Gamma_{15}$ levels consisting of the d$_{\rm xy}$, 
d$_{\rm yz}$, 
and $d_{\rm xz}$ and 2) the doubly degenerate $\Gamma_{12}$ consisting of 
d$_{\rm x^2-y^2}$ and d$_{\rm z^2}$.  The $\Gamma_{12}$ levels lie lower in 
energy.  It is important to note that in a tetrahedral 
field, only the $\Gamma_{15}$ states have the right symmetry to 
hybridize with the p-states of the sp neighbouring atom (As in this case), forming bonding and antibonding 
hybrid orbitals. The $\Gamma_{12}$ levels remain non-bonding and hence 
will be neglected in our hybridization analysis.  They will be assumed 
to constitute a full band (4 electrons).  The immediate problem with applying Hund's rule 
to the $\Gamma_{15}$ levels is that the effective moment on these levels 
is at least 2$\mu_B$ as found in recent calculations\cite{r1}. 
While inclusion of magnetic frustration\cite{si} might offer some
reduction in the moment, it offers no resolution of the problem that the 
moment lies in the xy plane.  The answer lies 
elsewhere as suggested by recent  first-principles calculations\cite{yild} and
a p-d mixing model\cite{tesanovic}.
 
The experimental observation that the Fe moment lies in the plane is 
highly suggestive of spin-orbit coupling.  To this end, our starting 
point is a general model, 
\beq 
H={\bf p}^2/(2m)+V_0+ \hbar/(4m^2c^2) \, (\nabla V_0\times {\bf p})\cdot\vec {\bf S} \, , 
\eeq 
for a cubic crystal with spin-orbit interaction 
where ${\bf p}$ is the momentum operator and ${\bf S}$ is the spin 
operator. This interaction breaks $SU(2)$ symmetry. The rough idea is to include the effects of p-d 
hybridization and the z-axis distortion through a series of 
successive diagonalizations to obtain the eigenstates in the final 
basis. We only include the outline of this calculation since an analogous 
analysis has been done for chalcopyrite 
semiconductors\cite{wolley}.  In obtaining the basis that 
diagonalizes the spin orbit interaction, we define 
$|\pm\rangle=|(X \pm iY)/\sqrt{2}\rangle$,$|0\rangle=|Z\rangle$ which are eigenstates of 
angular momentum $(L,L_z)$ with eigenvalues $(1,\pm 1)$ and $(1,0)$ 
respectively. The Hamiltonian for the p-levels is diagonalised through
\beq 
  \phi_{p1}^\alpha(\Gamma_8) &=& 1/\sqrt{3} |-\uparrow\rangle +\sqrt{2/3}|0\downarrow \rangle \ \ (J_z=-1/2), 
\nonumber 
\\ 
  \phi_{p0}^\alpha(\Gamma_8) &=& |+\uparrow\rangle \ \ (J_z=+3/2) , 
\nonumber 
\\ 
\phi_{p2}^\alpha(\Gamma_7) &=& \sqrt{2/3}|-\uparrow\rangle-1/\sqrt{3}|0\downarrow \rangle \ \ (J_z=-1/2), \\ 
\phi_{p1}^\beta(\Gamma_8) &=& -1/\sqrt{3}|+\downarrow\rangle+\sqrt{2/3}|0\uparrow \rangle \ \ (J_z=+1/2) , 
\nonumber 
\\ 
  \phi_{p0}^\beta(\Gamma_8) &=& |-\downarrow\rangle  \ \ (J_z=-3/2), \nonumber\\ 
  \phi_{p2}^\beta(\Gamma_7) &=& -\sqrt{2/3}|+\downarrow 
\rangle-1/\sqrt{3}|0\uparrow \rangle  \ \ (J_z=+1/2).\nonumber 
\eeq 
States with the same indices are degenerate and $\Gamma_n$ denotes the 
symmetry of a state.  Likewise, the basis for 
the d-levels which initially have $\Gamma_{15}$ symmetry, 
\beq 
  \phi_{d1}^\alpha(\Gamma_8) &=& 1/\sqrt{3}|\ominus\uparrow\rangle+\sqrt{2/3}|\odot\downarrow \rangle \ \ (J_z=-1/2), \nonumber\\ 
  \phi_{d0}^\alpha(\Gamma_8) &=& |\oplus\uparrow\rangle \ \ (J_z=+3/2) , \nonumber\\ 
  \phi_{d2}^\alpha(\Gamma_7) &=& \sqrt{2/3}|\ominus\uparrow\rangle-1/\sqrt{3}|\odot\downarrow \rangle \ \ (J_z=-1/2), \\ 
  \phi_{d1}^\beta(\Gamma_8) &=& -1/\sqrt{3}|\oplus\downarrow\rangle+\sqrt{2/3}|\odot\uparrow \rangle \ \ (J_z=+1/2) , \nonumber\\ 
  \phi_{d0}^\beta(\Gamma_8) &=& |\ominus\downarrow\rangle  \ \ 
  (J_z=-3/2), \nonumber\\ 
  \phi_{d2}^\beta(\Gamma_7) &=& -\sqrt{2/3}|\oplus\downarrow\rangle-1/\sqrt{3}|\odot\uparrow \rangle  \ \ (J_z=+1/2),\nonumber 
\eeq 
is formed from the states, $|\oplus\rangle=|(YZ+ i ZX)/\sqrt{2} \rangle$, 
$|\ominus\rangle=|(YZ- i ZX)/\sqrt{2} \rangle$, $| \odot \rangle=|XY\rangle$ are the 
eigenstates of $(L,L_z)$ with eigenvalues of $(2,+1)$,$(2,-1)$ and 
$(2,0)$, respectively.  As is clear, none of these states is an 
eigenstate of $S_z$ as is expected once the SU(2) spin symmetry is 
broken by the spin-orbit interaction. 
 
To consider the hybridization, we collate the states into two groups, 
segregating them according to their superscript $\alpha$ or 
$\beta$. Within each group they are ordered as follows: $\phi_{p1}^\alpha(\Gamma_8)$,$\phi_{d1}^\alpha(\Gamma_8)$, 
$\phi_{p0}^\alpha(\Gamma_8)$,$\phi_{d0}^\alpha(\Gamma_8)$, 
$\phi_{p2}^\alpha(\Gamma_7)$ and $\phi_{d2}^\alpha(\Gamma_7)$. Taking 
into consideration that only states of the same symmetry mix, 
\begin{equation} 
    \langle \phi_{pi}^\alpha(\Gamma_m)|V_{\rm pd}| 
    \phi_{dj}^\alpha(\Gamma_n)\rangle =M 
    \delta_{ij}\delta_{\alpha\beta}\delta_{mn}, 
\end{equation} 
we find that the hybridization matrix can be written as, 
\begin{eqnarray} 
%\begin{equation}\label{spandpd} 
V_{\rm pd}=\left[ 
         \begin{array}{cccccc} 
           \frac{\Delta_p}{3} & M & 0 & 0 & 0 & 0 \\ 
           M & E-\frac{\Delta_q}{3} & 0 & 0 & 0 & 0 \\ 
           0 & 0 & \frac{\Delta_p}{3} & M & 0 & 0 \\ 
           0 & 0 & M & E-\frac{\Delta_d}{3} & 0 & 0 \\ 
           0 & 0 & 0 & 0 & -2\frac{\Delta_p}{3} & M \\ 
           0 & 0 & 0 & 0 & M & E+2\frac{\Delta_d}{3} \\ 
         \end{array} 
       \right]. 
\end{eqnarray} 
$\Delta_p$ and $\Delta_q$ are the spin-orbit splitting of the p and 
d band, respectively.  The highest three eigen-energies are, 
\begin{eqnarray} 
  \lambda_0&=&\lambda_1 = \left[\Delta_p/3+E- 
\Delta_d/3\right]/2+  \sqrt{\alpha_1}/2, \\ 
\lambda_2 
&=&\left[-2\Delta_p/3+E+2\Delta_d/3\right]/2+ \sqrt{\alpha_2}/2 \, , 
\end{eqnarray} 
where $\alpha_1=(\Delta/3-E+\Delta_d/3)^2+4 M^2$ and 
$\alpha_2=(2\Delta/3 
+E+2\Delta_d/3)^2+4 M^2$. 
According to the symmetries $\Gamma_n$, $n=7$ or $n=8$, the corresponding eigenstates are 
   $ \Phi_i^\alpha(\Gamma_n)=a_i \phi_{pi}^\alpha(\Gamma_n)+b_i 
    \phi_{di}^\alpha(\Gamma_n)$ 
with the coefficients $a_i$ ($i=0,1,2$) and  $b_i$ defined as, 
\begin{eqnarray} 
\gamma_0=\gamma_1=a_1^2=1-b_1^2=\left[1+M^2/\left(\lambda_1-E+\Delta_d/3\right)^2\right]^{-1}, 
\nonumber\\ 
\gamma_2=a_2^2=1-b_2^2=\left[1+M^2/\left(\lambda_2-E-2\Delta_d/3\right)^2\right]^{-1}.\nonumber 
\end{eqnarray} 
To gain information about the spins, we transform the operator for the 
z-projection of the spin, 
\begin{eqnarray} 
  \sigma_z^{{\rm I}_\alpha} &=& \frac{1}{3} \left[ 
                            \begin{array}{cccccc} 
                              -1 & 0 & 0 & 0 & 2\sqrt{2} & 0 \\ 
                              0 & -1 & 0 & 0 & 0 & 2\sqrt{2} \\ 
                              0 & 0 & 3 & 0 & 0 & 0 \\ 
                              0 & 0 & 0 & 3 & 0 & 0 \\ 
                              2\sqrt{2} & 0 & 0 & 0 & 1 & 0 \\ 
                              0 & 2\sqrt{2} & 0 & 0 & 0 & 1 \\ 
                            \end{array} 
                          \right] \, , 
\end{eqnarray} 
and $\sigma_z^{{\rm I}_\beta} = -\sigma_z^{{\rm I}_\alpha}$ 
into the ${\rm I}_\alpha$ or ${\rm I}_\beta$ basis. 
 
Ultimately, we will focus only on the three highest eigenstates.  We 
refer to this reduced basis as ${\rm II}^{\alpha,\beta}$. The final 
ingredient is the z-axis distortion from perfect cubic symmetry. We 
consider a crystal field interaction of the form, 
\begin{eqnarray} 
  \langle X|V_{\rm cf}| X\rangle &=& \zeta\langle Y|V_{\rm cf}| Y\rangle =\delta_p/3, 
\langle Z|V_{\rm cf}| Z\rangle =- 2\delta_p/3 \, , 
\nonumber\\ 
   \langle ZX|V_{\rm cf}| ZX\rangle &=& \zeta\langle YZ|V_{\rm cf}| YZ\rangle 
   =\delta_d/3 \, , 
\nonumber\\ 
\langle XY|V_{\rm cf}| 
   XY\rangle &=&-2\delta_p/3.\nonumber 
\end{eqnarray} 
The parameter $\zeta$ accounts for the distortion in the $a-b$ plane. Experimentally, the lattice constants along $a$ and $b$ differ by $0.3\%$. While any distortion is sufficient to lower the $U(1)$ rotational symmetry in the plane to simply $Z_2$ (Ising), this effect is small relative to the overall z-axis tilt.  As the parameter $\delta_p$ is certainly not known within $0.3\%$, we consider onlythe case of $\zeta=1$. 
The crystal field Hamiltonian in the ${\rm II}_{R}^{\alpha,\beta}$ 
is, 
\begin{equation} 
    V_{\rm cf}=\left[ 
           \begin{array}{ccc} 
             \lambda_1-\Gamma_1 & 0 &\Gamma_2\\ 
             0 & \lambda_0+\Gamma_1 & 0 \\ 
             \Gamma_2 & 0 & \lambda_2 \\ 
           \end{array} 
         \right] \, , 
\end{equation} 
where $\Gamma_1=\frac{1}{3}[\delta_p\gamma_1+\delta_d(1-\gamma_1)]$ 
and 
$\Gamma_2=\frac{\sqrt{2}}{3}[\delta_p\sqrt{\gamma_1\gamma_2}+\delta_d\sqrt{(1-\gamma_1)(1-\gamma_2)}]$. 
Diagonalizing this Hamiltonian gives rise the following three 
energy levels,  $E_0(\Gamma_6) =\lambda_0+\Gamma_1$ and 
\begin{eqnarray} 
  2 E_{1,2} (\Gamma_7)&=& \lambda_1+\lambda_2-\Gamma_1 \pm 
  \sqrt{\left(\lambda_1-\lambda_2-\Gamma_1\right)^2+4 
  \Gamma_2^2}, \nonumber. 
\end{eqnarray} 
and their corresponding eigenstates, 
\begin{eqnarray} 
\Psi_0^{\alpha,\beta}(\Gamma_6)&=&\Phi_0^{\alpha,\beta}(\Gamma_7),  \\ 
    \Psi_i^{\alpha,\beta}(\Gamma_7)&=&c_1 
    \Phi_i^{\alpha,\beta}(\Gamma_8)+d_1\Phi_i^{\alpha,\beta}(\Gamma_7), 
\end{eqnarray} 
which we refer to as the ${\rm III}^{\alpha(\beta)}$ basis, where 
$i=1,2$ and $c_i$ 
and $d_i$ are defined as, $c_1^2 =1-d_1^2=1/[1+\Gamma_2^2/(E_1-\lambda_2)]$ 
with $d_2 = -c_1$, and $c_2=d_1$. 
In the final basis, ${\rm III}^{\alpha(\beta)}$, the $\sigma_z$ spin 
matrix becomes\begin{equation}\label{spinf} 
    \sigma_z^{{\rm III}_R^{\alpha}}=- \sigma_z^{{\rm III}_R^{\beta}}=\left[ 
           \begin{array}{ccc} 
             e & 0 & f \\ 
             0 & 1 & 0 \\ 
             f & 0 &  -e\\ 
           \end{array} 
         \right] \, , 
\end{equation} 
where $e=2c_1d_1 
\Gamma+(d_1^2-c_1^2)/3$, $f=(d_1^2-c_1^2)\Gamma-2c_1d_1/3$ 
and $\Gamma=2\sqrt{2}(a_1 a_2+b_1 b_2)/3$. So we can see 
clearly that the final basis does not diagonalise the the spin 
matrix.  Consequently, the states ${\rm III}^{\alpha(\beta)}$ 
represent some linear combination of spin up and spin down.  A 
crucial point about this spin matrix: the $\alpha$ and $\beta$ 
states have opposite projections of spin in the states with energies 
$E_1$ and $E_2$. 
 
In the transformed basis, the resultant Hamiltonian reads, 
 \beq\label{hfinal}  
H &=& \sum_{a,\mu,i} E_{a} c^{\dag}_{a,\mu,i} c_{a,\mu,i} 
-\!\! \sum_{a,b,\mu,\nu, \langle i,j \rangle} t_{a,b}^{\mu,\nu} 
c^{\dag}_{a,\mu,i} c_{b,\nu,j} + {\rm h.c.}\nonumber\\ 
&+&\sum_{a,b\mu,\nu,i} 
U_{ab}^{\mu\nu} n_{i a\mu}n_{i b\nu} \, , 
\end{eqnarray} 
where $a,b =0,1,2$ and $\mu,\nu=\alpha,\beta$, and
\begin{eqnarray} 
  t_0&=& t_{00}^{\alpha\alpha} =t_{00}^{\beta\beta}= a_1^4 t_p+b_1^4 t_d  , 
\nonumber\\ 
  t_1 &=&t_{11}^{\alpha\alpha}=t_{11}^{\beta\beta}= (c_1^2a_1^2+d_1^2a_2^2)t_p+(c_1^2b_1^2+d_1^2b_2^2)t_d \, , 
\nonumber \\ 
  t_2 &=&t_{22}^{\alpha\alpha}=t_{22}^{\beta\beta}= (d_1^2a_1^2+c_1^2a_2^2)t_p+(d_1^2b_1^2+c_1^2b_2^2)t_d\, , 
\nonumber \\ 
  t_{12}&=& t_{12}^{\alpha\beta}= c_1d_1(a_1^2-a_2^2)t_p+c_1 
  d_1(b_1^2-b_2^2)t_d, 
\nonumber\\ 
  U_0^{\alpha\beta}&\equiv& U_{00}^{\alpha\beta}=a_1^4 U_p/2+b_1^4 U_d/2,\nonumber\\ 
  U_i^{\alpha\beta} &=&(C_i^4+A_i^4/2) U_p+(D_i^4+B_i^4/2) U_d,\ \ (i=1,2) 
\nonumber\\ 
  U_{12}^{\alpha\beta}&=&[(C_1C_2)^2+A_1 A_2)^2/2] U_p+[(D_1D_2)^2+(B_1 B_2)^2/2] U_d , 
\nonumber\\ 
  U_{0i}^{\alpha\beta}&=&(a_1A_i)^2 U_p/2+(b_1 B_i)^2 U_d/2 \ \ (i=1,2) 
 %U_{02}^{\alpha\beta}&=&(a_1A_2)^2U_p+(b_1 B_2)^2 U_d. 
\end{eqnarray} 
All other couplings, for example, $t_{01}$ and $t_{02}$ vanish by 
symmetry.  The coefficients $A_i$,$B_i$,$C_i$ and $D_i$ are defined as, 
\begin{eqnarray} 
  A_i &=& 1/\sqrt{3}c_i a_1+\sqrt{2/3}d_i 
  a_1 , \ 
B_i = 1/\sqrt{3}c_i b_1+\sqrt{2/3}d_i 
  b_1 ,\nonumber\\ 
  C_i &=& \sqrt{2/3}c_i a_1-1/\sqrt{3}d_i 
  a_1 , \ 
 D_i = \sqrt{2/3}c_i b_1-1/\sqrt{3}d_i 
  b_1 , \nonumber 
\end{eqnarray} 
where $c_1$, $d_1$, $a_i$, $b_i$ are defined as before. 
 
\begin{table}[!h] 
\tabcolsep 0pt \caption{Energy levels, hopping matrix elements, and 
 interactions in the three 
highest levels in transformed basis, ${\rm III}^{\alpha(\beta)}$.} 
\vspace*{-24pt} 
\begin{center} 
\def\temptablewidth{0.5\textwidth} 
{\rule{\temptablewidth}{1pt}} 
\begin{tabular*}{\temptablewidth}{@{\extracolsep{\fill}}|c|c|c|c|c|c|c|c|c|c|c|c|} 
\hline  $\delta_{p(d)}$ & $E_0$ &$E_1$ & $E_2$ & $t_{0(1,2)}  $ & 
$t_{12}^{\alpha\beta}$& $U_0$ &$U_1$&$U_2$ & $U_{12}$ &$U_{01}$ 
&$U_{02}$ 
  \\ \hline $0.06$ & $2.21$ & $2.22$ &$2.15$ & $0.7$ & $0$ & $1.54$ &$1.55$ &
$2.98$ & $0.12$ &$1.54$ &$0.035$

  \\ \hline

  $0$ & $2.22$ &$2.21$&$2.19$ & $0.7$ & $0$ & $1.54$ &$1.11$& $1.54$
&

  $1.07$ &$1.07$ &$0.51$\\

       \end{tabular*} 
       {\rule{\temptablewidth}{1pt}} 
       \end{center} 
       \end{table} 
 
\begin{figure} 
  \includegraphics[width=9cm]{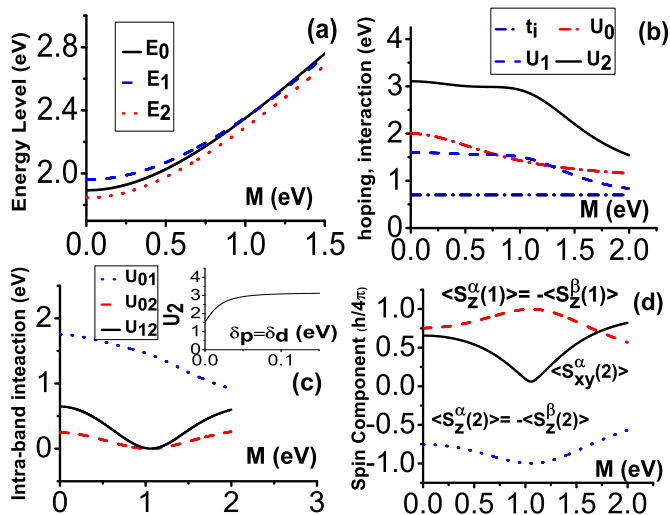}\\ 
  \caption{Energy levels, hoping matrix elements, on-site interactions, intra-band interaction and spin components as 
  a function of hybridization $M$. Shown in the inset of (c) is the 
  on-site 
interaction $U_2$ as a function of the monoclinic distortion. }\label{fig2} 
\end{figure}

We analyzed the energy levels, interaction strengths, and spin 
projections based on representative values for iron-based systems. 
For instance, if we set
$U_p=U_d=4eV$\cite{haule}, $t_p=t_d=0.7 eV$\cite{r1,haule}, 
$M=0.8 eV$\cite{r1,si}, $\Delta_p=0.426 eV$\cite{Herman}, 
$\Delta_d=0.08 eV$\cite{cheli},  $E=1.9 eV$\cite{r1}, and 
$\delta_p=\delta_d=0.06 eV$\cite{Shorikov} we arrive at the
parameters of Table I for Hamiltonian (\ref{hfinal}). Notice,
however, that there is an uncertainty in the value of these parameters,
especially the hybridization energy, $M$ and the monoclinic distortion. Hence, we explore the
dependence of the Hamiltonian parameters on both.
 As shown in the first panel of Fig.~\ref{fig2}(a), 
the $E_2$ level is the lowest followed by 
$E_0$ and then $E_1$ for sufficiently large values of $M$. As can be seen from Fig.~\ref{fig2}(b), 
$U_2>E_0-E_2$ and $E_1-E_2$. Consequently, $E_2$ and one of $E_0$ or $E_1$ will be 
singly occupied. To determine the ground state spin configuration, 
we note that both the interactions $U_{12}^{\alpha\beta}\ne 0$ and
$U_{02}^{\alpha\beta}\ne 0$ (see Fig. (\ref{fig2}c)) with 
$\alpha\ne\beta$ for the highest two occupied levels.  Recall, the 
$z$-projection of the spins in $\alpha$ and $\beta$ is reversed in 
levels $E_0$ (or $E_1$) relative to $E_2$. Level $E_0$ is an
eigenstate of $S_z$ while $E_1$ is not.  Consequently, the lowest energy 
configuration for single occupancy of the levels $E_0$ and $E_2$ is 
an antiparallel alignment of the spins. That is, both electrons are 
in the $\alpha$ or in the $\beta$ state in both levels.  This 
configuration does not cost the repulsion term $U_{02}^{\alpha\beta}$
or $U_{12}^{\alpha\beta}$. 
Since there are 12 electrons to fill these levels, we arrive at 
the filling structure shown in Fig.~\ref{fig1}.  Consequently, the 
three effects considered preclude a naive assignment of the spins 
according to Hund's rule\cite{tesanovic}.  This conclusion is a general result of 
this analysis, not an artifact of fine-tuning the bare parameters. 
 
The problem has now been reduced to the physics of two low-lying energy
levels, $E_1$ and $E_2$. That a two-band reduction reproduces\cite{scalapino} the Fermi surface seen in the local density approximation\cite{lda} and experiments\cite{osc} corroborates our approach. If we neglect the interactions in  
(\ref{hfinal}), the problem can be easily diagonalized and one finds two,
doubly degenerate, energy-shifted bands. The splitting between the bands is due to the crystal
field that shifts each half-filled band away from 
the perfect nesting condition. As a result, it is difficult to reconcile the 
experimental observation of a spin-density wave (SDW) with a simple 
Fermi surface instability due to nesting. The resolution may lie in 
the interactions. As Fig.~\ref{fig2}(b) indicates, for all values of the hybridization, the 
levels 1 and 2 have differing p and d character.  In $E_2$ the 
on-site interaction, $U_2$ exceeds $U_0$ (or $U_1$) by more than a factor of two 
at $M=0.8eV$. 
Consequently, single occupancy in the $E_0$ (or the $E_1$) level results in itineracy 
whereas in the $E_2$ level Mott physics can be relevant since $U_2\approx 4 
t_2$.  This difference is due entirely to the different p-d character 
between the $E_0$ (or $E_1$) and $E_2$ levels which is expected as they are orthogonal. Finally, we show in the last panel in Fig.~\ref{fig2}(d) 
the value of $S_z$ in the levels $E_1$ and $E_2$. Recall level $E_0$ is an eigenstate of $S_z$ with the $z-$projection opposite to that in
$E_2$.  As shown in the last panel in Fig. (\ref{fig2}), the value of
$S_z$ in $E_2$ is $0.95\mu_B$.  If $E_0$ is the next lowest level,
a net moment in the $z-$direction of $0.06\mu_B$ remains as has been recently observed\cite{o2}.  If $E_1$ is
relevant (as would be the case for  $M>0.83eV$), the
z-moment vanishes as shown in Fig. (\ref{fig2}d). The itineracy of the electrons in the $E_0$ level does not affect this cancellation as it is the average not the local spin configuration that is relevant in a magnetization measurement. Hence, for experimentally relevant values of $M$ ($.5eV<M<1.0eV$), the residual z-component of the moment is strongly diminished.
The remaining x-y component on $E_2$ is  $S_{xy}=\sqrt{1-\langle S_z 
  \rangle^2}=0.317\mu_B$.   Such a moment can order via the
super-exchange mechanism on $E_2$ as $U_2\gg t_2$. The evolution of $S_{\rm xy}$ as a function 
of $M$ in Fig.~\ref{fig2}(d) shows that the analysis here is consistent 
with the range of the magnetic moment seen experimentally\cite{lynn,o5}. 
 
Our analysis also explains why the structural 
transition\cite{lynn} at 150K must precede the onset of antiferromagnetism. 
As the inset in Fig.~\ref{fig2}(c) indicates, the on-site energy $U_2$ 
diminishes as the crystal field associated with the monoclinic 
distortion vanishes.  Once $U_2<t_2$, a transition to an 
antiferromagnet via the super-exchange interaction is untenable. The 
structural transition breaks rotational symmetry in the plane not SU(2) 
which is already broken at the outset by spin orbit interaction. The 
success of the analysis presented here in describing the 
antiferromagnet in the parent material, LaFeAsO, implies that Eq. (\ref{hfinal}) should be used in 
any subsequent analysis of superconductivity.  The presence of an 
itinerant band coupled to one with moderate Mott 
physics makes the problem of the iron pnictides more akin\cite{brink} to that of 
the Kondo lattice in heavy fermions than the cuprates.
 
P.P. was supported in part by the NSF 
DMR-0605769.  P. P. thanks his students Weicheng Lv for 
double-checking the calculations and T.-P. Choy for his 
characteristically level-headed remarks. A.~H.~C.~N. thanks 
A. Polkovnikov for pointing out this problem to him. 

{\it Note Added} While this work was under review, McGuire, et al\cite{o2} reported a residual magnetic moment along the c-axis equal to 0.06$\mu_B$ as predicted here.

\end{document}